\documentclass[pra,aps,groupedaddress,floatfix,twocolumn,superscriptaddress,showpacs]{revtex4}
\usepackage{amsmath,amssymb,multirow,epsfig,bm,pifont}
\usepackage[linkcolor=black,citecolor=black,urlcolor=blue,colorlinks=true,linktocpage=true]{hyperref}
\usepackage{graphicx}
\usepackage{epsfig}
\usepackage{bm}
\usepackage{tabularx}
\usepackage{makecell}

\newcommand{\beq}{\begin{equation}}
\newcommand{\eeq}{\end{equation}}
\newcommand{\bea}{\begin{eqnarray}}
\newcommand{\eea}{\end{eqnarray}}

\begin{document}

%%%%%%%%%%%%%%%%%%%%%%%%%%%%%%%%%%%%%%%%%%%%%%%%%%%%%%%%%%%
\title{Harmonically trapped fermions in two dimensions: ground-state energy 
and contact of SU(2) and SU(4) systems via nonuniform lattice Monte Carlo}

\author{Z.-H. Luo}
\affiliation{Department of Applied Physics, South China Agricultural University, Guangzhou, 510642, China}
\affiliation{Department of Physics and Astronomy, University of North Carolina, Chapel Hill, North Carolina 27599-3255, USA}

\author{C. E. Berger}
\affiliation{Department of Physics and Astronomy, University of North Carolina, Chapel Hill, North Carolina 27599-3255, USA}

\author{J. E. Drut}
\affiliation{Department of Physics and Astronomy, University of North Carolina, Chapel Hill, North Carolina 27599-3255, USA}

\begin{abstract}
We study harmonically trapped, unpolarized fermion systems with attractive interactions in two spatial dimensions
with spin degeneracies $N_f^{} = 2$ and $4$ and $N/N_f^{} = 1, 3, 5$ and 7 particles per flavor.
We carry out our calculations using our recently proposed quantum Monte Carlo method on a nonuniform lattice. 
We report on the ground-state energy and contact for a range of couplings, as determined by the binding energy of the
two-body system, and show explicitly how the physics of the $N_f$-body sector dominates as the coupling is increased.
\end{abstract}

\date{\today}


\maketitle

%%%%%%%%%%%%%%%%%%%%%%%%%%%%%%%%%%%%%%%%%%%%%
% Introduction
\section{Introduction}

In the past year, there have been multiple reports on experiments with ultracold fermionic atoms 
in constrained, quasi-two-dimensional optical traps. For instance, the Berezinskii-Kosterlitz-Thouless (BKT) 
superfluid transition~\cite{BKT} was observed in Refs.~\cite{Experiments2D2015PairCondensation,Experiments2D2015BKTObservation}, and the
finite-temperature thermodynamics was studied in Refs.~\cite{ValeThermo, HeidelbergThermo}.
The first realization of 2D systems was in fact reported only a few years ago 
in~\cite{Experiments2D2010Observation, Experiments2D2011Observation}, and since then
multiple efforts followed, such as radiofrequency spectroscopy~\cite{Experiments2D2011RfSpectroscopy, Experiments2D2012RfSpectraMolecules},
studies of dimensional crossover~\cite{Experiments2D2011Crossover2D3D, Experiments2D2012Crossover2D3D},
polarons~\cite{Experiments2D2012Polarons}, density distribution~\cite{Experiments2D2011DensityDistributionTrapped}, 
viscosity~\cite{Experiments2D2012Viscosity}, Tan's contact~\cite{ContactExperiment2D2012}, ground-state pressure~\cite{Experiments2D2014},
and polarized systems~\cite{Experiments2D2015SpinImbalancedGas}. (See also~\cite{RanderiaPairingFlatLand,PieriDanceInDisk}).

Experiments continue to move forward at an exciting pace, and consistent advances are seen on the theory side as well.
Early analytic studies considered pairing in the 2D Bose-Einstein condensation 
(BEC) and Bardeen-Cooper-Schrieffer (BCS) crossover at the mean-field level~\cite{Miyake,BCSBEC2D,ZhangLinDuan}. 
The ground-state equation of state was obtained in an ab initio fashion only in 2011, in Ref.~\cite{Bertaina}.
Reference~\cite{ShiChiesaZhang} followed up with a more detailed first-principles study of the ground state 
where the pressure, contact, pairing properties, and condensate fraction were determined.
The thermal equation of state was first computed in Ref.~\cite{LiuHuDrummond} in the virial expansion,
and in the Luttinger-Ward approach in Ref.~\cite{Enss2D}.
Pair correlations were investigated in Refs.~\cite{ParishEtAl} in dilute, high-temperature regimes using the virial expansion,
and in Ref.~\cite{BarthHofmann}, which also analyzed Tan's contact.
The work of Refs.~\cite{ChaffinSchaefer,EnssUrban,BaurVogt} studied collective modes, while the shear viscosity and 
spin diffusion were calculated in Ref.~\cite{EnssShear}. Finite-temperature quantum Monte Carlo calculations characterized the
density, pressure, compressibility, and contact more recently in Ref.~\cite{AndersonDrut}, and a comparison
between theory and experiment was carried out in~\cite{HuVale}.

The present work aims to complement some of the above computational studies by reporting our Monte Carlo
calculations of the ground-state energy and contact of 2D fermions in a harmonic trap. Our calculations
were performed in a non-uniform lattice, a technique put forward in Ref.~\cite{GHBDA}.
We study spin degeneracies $N_f^{}=2$ and $4$ and unpolarized systems of $N$ particles for $N/N_f^{} = 1,3,5,7$.
In this first paper we do not study higher values of $N_f^{}$, although such calculations are certainly feasible with 
the same methods. This is particularly interesting given the progress in the experimental realization of SU($N_f^{}$)-symmetric systems 
in the last few years, in particular in the presence of optical lattices~\cite{CazalillaRey}. Moreover,
experiments involving a small number of atoms have been achieved as well~\cite{HeidelbergFewBodyExp}, and 
for those experiments, if ever carried out in 2D, the present work represents a prediction (see~\cite{ParishReview} for
a recent review).

%%%%%%%%%%%%%%%%%%%%%%%%%%%%%%%%%%%%%%%%%%%%%%%%%%%%%%%%%%%
\section{Hamiltonian and many-body method}

As mentioned above, we focus here on a two-dimensional system of $N^{}_f$ fermion species, attractively interacting via pairwise 
interactions. The full Hamiltonian in second quantization form is
\beq
\hat H = \hat T + \hat V^{}_\text{ext} + \hat V^{}_\text{int},
\eeq
where 
\beq
\hat T = \sum_{s=1}^{N^{}_f}\int\! d^2 p\; \left(\frac{{\bf p}^2}{2m}\right)\; \hat n_{s}^{}({\bf p})
\eeq
is the kinetic energy operator,
\beq
\hat V^{}_\text{ext} = \sum_{s=1}^{N^{}_f}\int\! d^2x\, \left(\frac{1}{2} m \omega^2 {\bf x}^2 \right) \hat n_{s}^{}({\bf x})
\eeq
is the external potential energy operator, and
\beq
\hat V^{}_\text{int} = -\frac{g}{2} \sum_{s \neq s'} \int\! d^2x\, \hat n_{s}^{}({\bf x}) \hat n_{s'}^{}({\bf x})
\eeq
is the two-body interaction operator.
In the above equations, 
$\hat n_{s}^{}({\bf p})$ and $\hat n_{s}^{}({\bf x})$ are, respectively, the particle-density operators in coordinate and momentum 
space for species $s$, and we have included an overall factor of $1/2$ to avoid over-counting in the flavor sum.

As in Ref.~\cite{GHBDA}, but now in 2D, we place the system in a discretized space of $N^{}_x \times N^{}_x$ points using the 
Gauss-Hermite (GH) lattice $\{x_i^{},y_j^{}\}$ and weights $\{w_i^{},w_j^{}\}$ of gaussian quadratures in each direction to define such a lattice~\cite{NR}.
The discretized form of the interaction then becomes
\beq
\hat V^{}_\text{int} = -\,\frac{g}{2} \sum_{s \neq s'} \sum_{i,j=1}^{N_x^{}} {w}_i^{}{w}_j^{} e^{x_i^2 + y_j^2} \ \hat n_{s, (i,j)}^{} \, \hat n_{s', (i,j)}^{},
\eeq
where $\hat n_{s, (i, j)}^{}$ is the lattice density operator for spin $s$ at position $(i,j)$.
Thus, we obtain a position-dependent coupling constant $g(x^{}_i,y^{}_j) = g \, {w}_i^{}{w}_j^{} e^{x_i^2 + y_j^2}$.

Following the usual path of the lattice Monte Carlo formalism, we then approximate 
the Boltzmann weight using a symmetric Suzuki-Trotter decomposition: 
\beq
\label{Eq:TS}
e^{-\tau\hat H} = e^{-\tau/2 (\hat T + \hat V^{}_\text{ext})}e^{-\tau \hat V^{}_\text{int}}e^{-\tau/2 (\hat T + \hat V^{}_\text{ext})}
+\mathcal O(\tau^3),
\eeq
for some small temporal discretization parameter $\tau$ (which below we take to be $\tau = 0.05$ in lattice units). 
This discretization of imaginary time results in a temporal
lattice of extent $N_\tau^{}$, which we also refer to below in terms of $\beta = \tau N_\tau^{}$ and in dimensionless 
form as $\beta \omega$. A Hubbard-Stratonovich (HS) transformation~\cite{HS} of the interaction factor is then used to
represent the interaction using an auxiliary-field (see e.g.~\cite{MCReviews}), which results in a field-integral form of 
the left-hand side of Eq.~\ref{Eq:TS}. We use that form combined with the power-projection method~\cite{GolubVanLoan} 
to obtain ground-state properties of the system, using a Slater determinant of harmonic oscillator (HO) single-particle orbitals
as a trial wavefunction.

\begin{figure}[t]
\includegraphics[width=1.0\columnwidth]{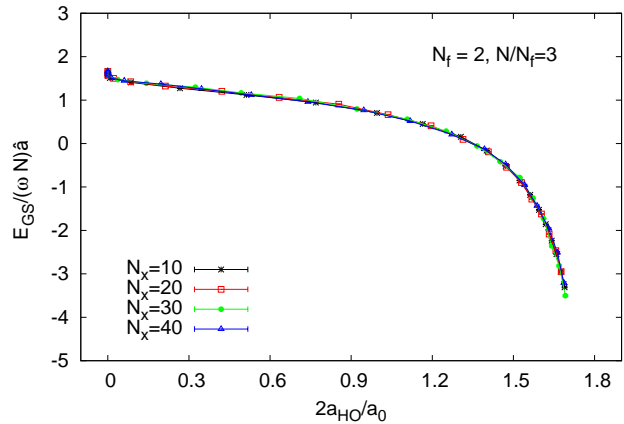}
\caption{\label{Fig:2DN6Energy}(color online)
Ground-state energy per particle of six harmonically trapped spin-$1/2$ fermions in 2D 
as a function of the coupling strength $2 a_\text{HO}/a_0$ for 4 different lattice sizes $N_x^{}=$10, 20, 30, 40. 
The error bars reflect the statistical uncertainty. The exact result at $2 a_\text{HO}/a_0=0$, i.e. for the noninteracting
case, is $E/(\hbar \omega N) = 5/3$, but the approach is logarithmic, which explains the peak-like structure at very
weak coupling (see figures below).
}
\end{figure}

As in our previous work, we tune the system to specific physical points by way of the 
2D scattering length $a^{}_{0}$, which we present everywhere in units of the HO length scale 
$a^{}_\text{HO}$ (which is $1$ in our units). To this end, we computed 
the ground-state energy $E^{}_\text{GS}$ of the two-body problem and matched it to that of the continuum solution, for which
the relationship between $E^{}_\text{GS}$ and the scattering length is well known (see, e.g., Ref.~\cite{BuschEtAl}).
We used this renormalization procedure for each lattice size, and then proceeded to higher particle numbers using the coupling thus
determined. To illustrate the success of the procedure, we show the results for the unpolarized spin-$1/2$ six-body problem in 
Fig.~\ref{Fig:2DN6Energy} for several lattice sizes. As can be appreciated in that figure, the finite-size effects are vanishingly small
for $N_f^{}=2$.

%%%%%%%%%%%%%%%%%%%%%%%%%%%%%%%%%%%%%%%%%%%%%%%%%%%%%%%%%%%
\section{Analysis and Results}

In this section we present our results for the energy per particle and Tan's contact. In all of our tests, as illustrated in 
the previous section, the lattice-size effects were very small. However, increasing $N_f^{}$ effectively enhances the
attractive interaction, such that bound states become even more deeply bound, which in turn amplifies lattice-spacing 
effects (see also Ref.~\cite{DeanLeeClusters2D}).
For this reason, we do not consider $N_f^{} > 4$ in this work. As a compromise with the
computational cost of running the calculations, we chose to fix $N_x^{}=10$ throughout and explore a range of
values of $N/N^{}_f$. To minimize statistical effects, we took $10^4$ decorrelated samples of the auxiliary field 
$\sigma$, which results in a statistical uncertainty of order $1\%$.

%%%%%%%%%%%%%%%%%%%%%%%
\subsection{Ground-state energy}

\begin{figure}[t]
\includegraphics[width=1.0\columnwidth]{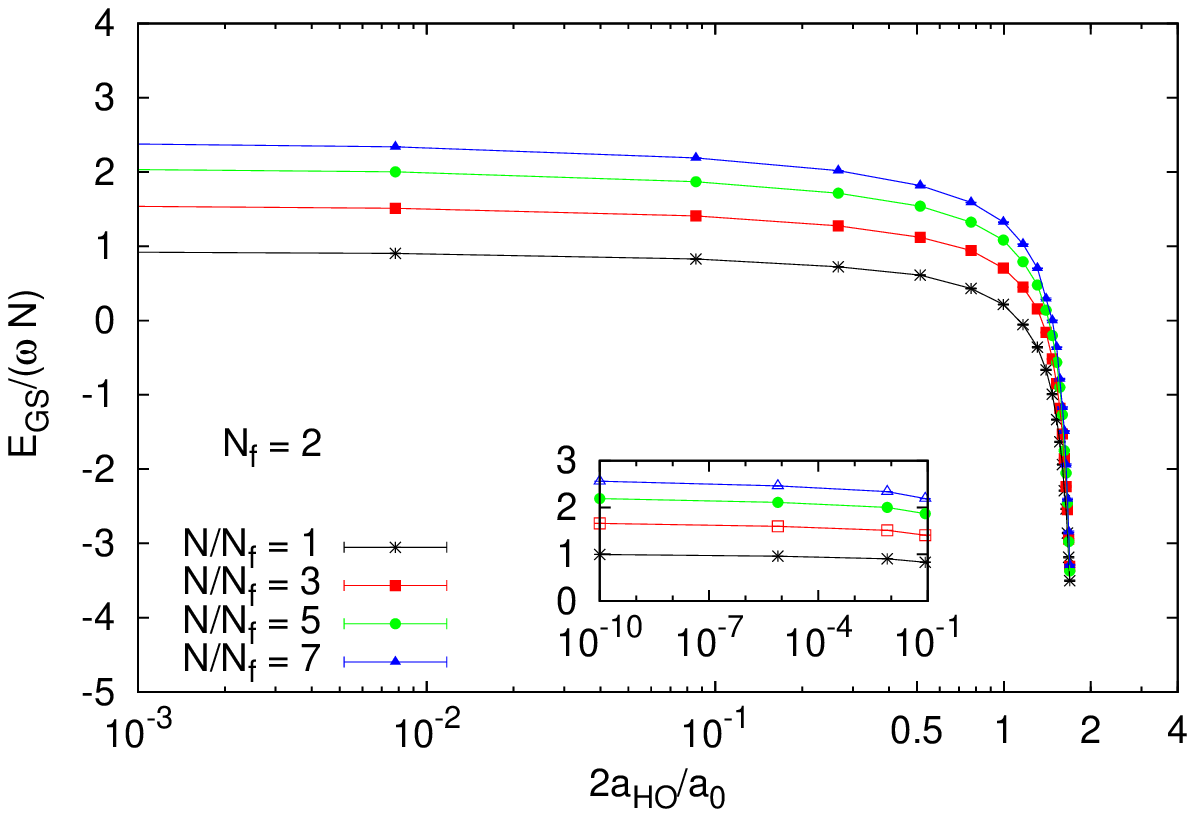}
\includegraphics[width=1.0\columnwidth]{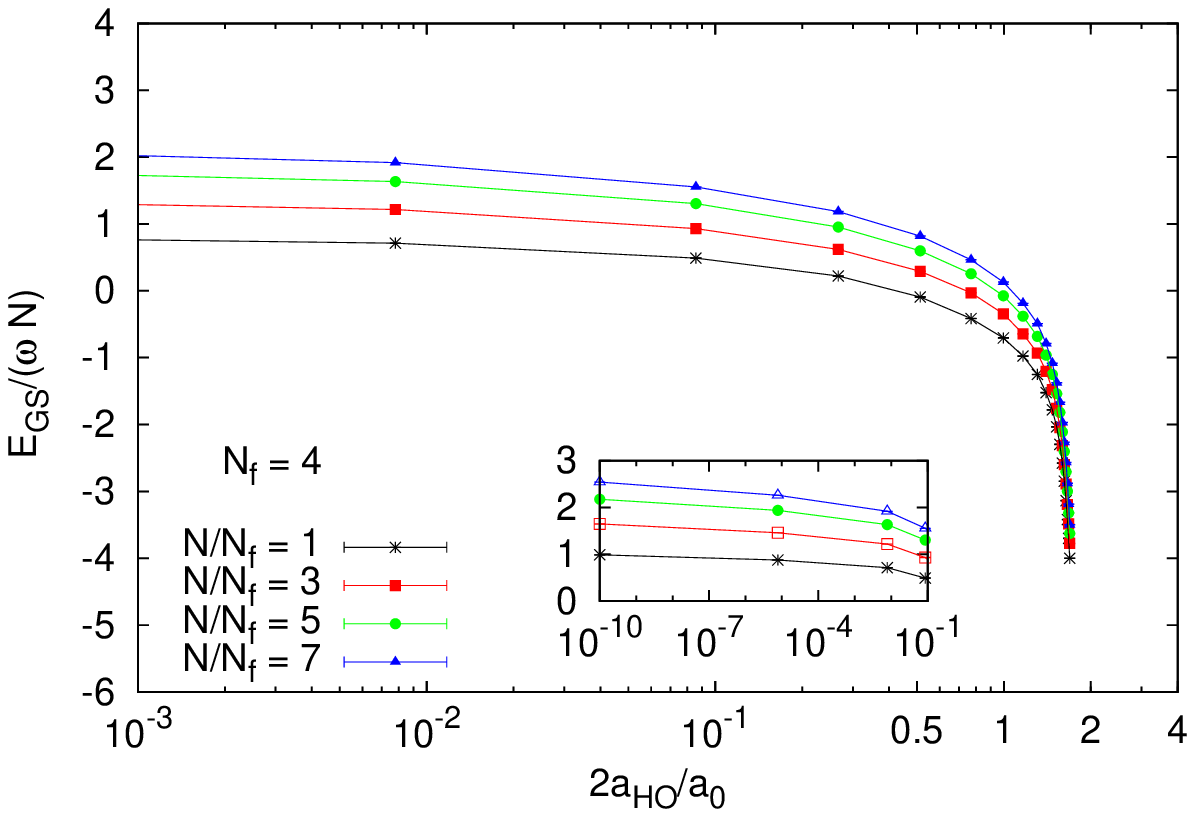}
\caption{\label{Fig:2D_energy_Nf2_Nx10}(color online)
Ground-state energy of $N^{}_f=2$ (top) and 4 (bottom) species of harmonically trapped fermions in 2D 
as a function of the coupling strength $2 a_\text{HO}/a_0$, for particle numbers $N/N^{}_f=$1, 3, 5, 7 (from bottom to top).
The error bars reflect the statistical uncertainty. The inset shows the (logarithmic) approach to the non-interacting limit. 
For $N^{}_f=2$ the exact values of $E^{}_\text{GS}/N$ in the noninteracting limit are (from bottom to top): 1, 5/3, 12/5, 3.
}
\end{figure}

\begin{figure}[h]
\includegraphics[width=1.0\columnwidth]{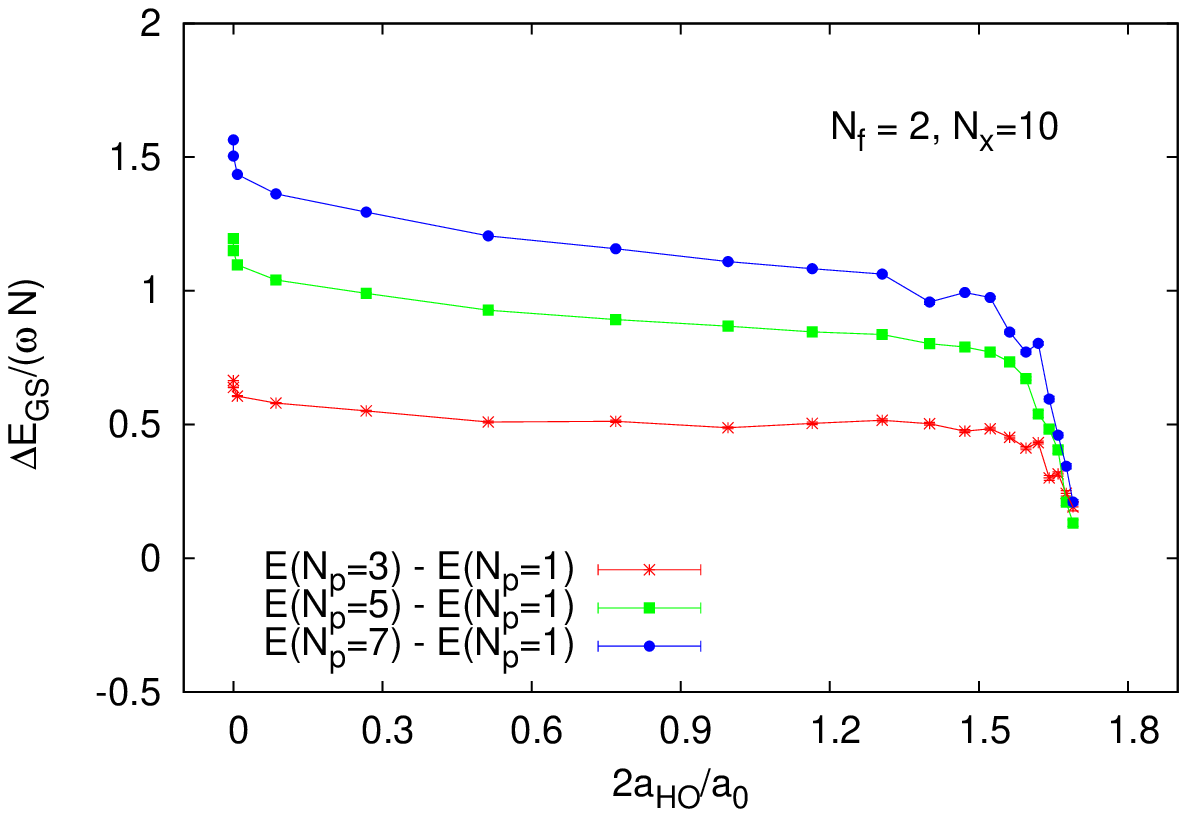}
\includegraphics[width=1.0\columnwidth]{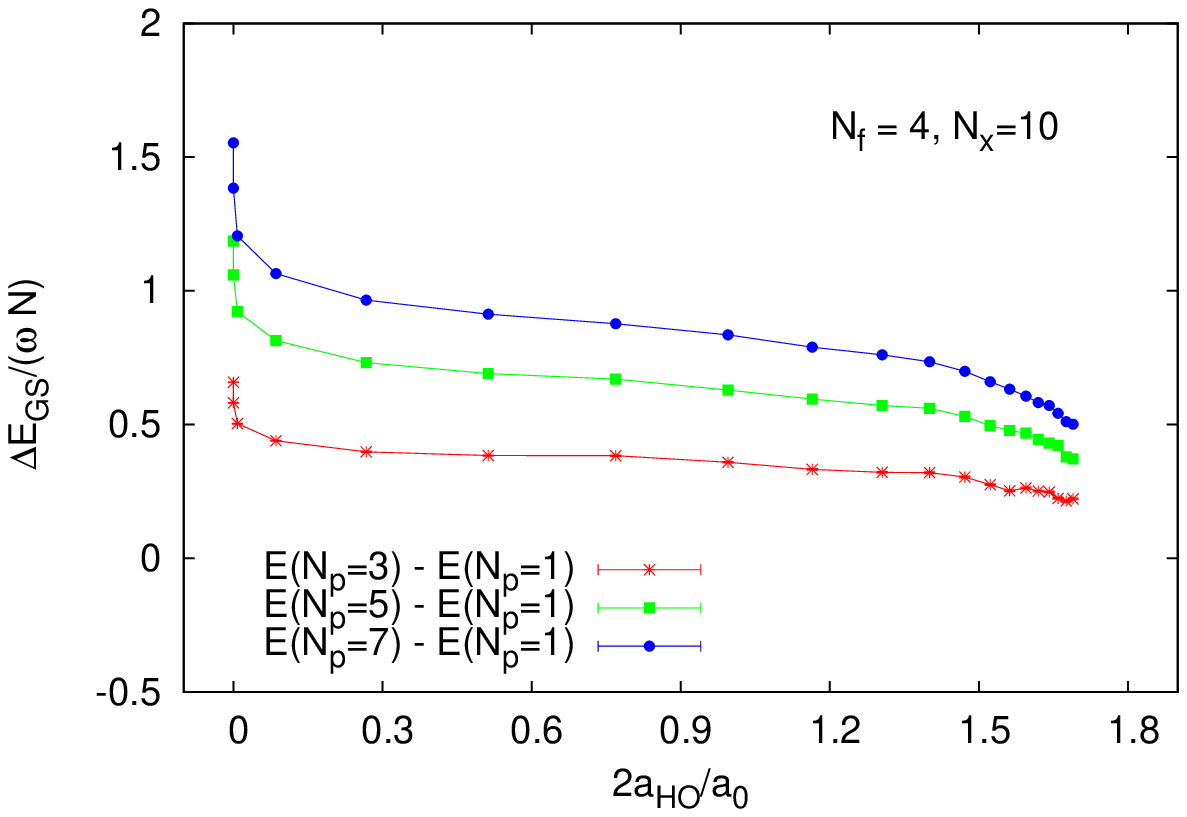}
\caption{\label{Fig:2D_energy_Nf4_Nx10Diff}(color online) Energy per particle difference, taking $N/N^{}_f=$1 as 
a reference, for $N^{}_f=2$ and $N^{}_f=4$ species of harmonically trapped fermions in 2D as a function of the coupling strength 
$2 a_\text{HO}/a_0$, and for particle numbers $N/N^{}_f=$3, 5, 7 (from bottom to top).
}
\end{figure}

In Fig.~\ref{Fig:2D_energy_Nf2_Nx10} we show our results for the ground-state energy of systems with $N^{}_f$ = 2 and 4.
For each flavor number, we studied systems with $N/N^{}_f$ = 1, 3, 5, and 7 particles per flavor. In all cases, as evident from the figures,
the energy {\it per particle} monotonically increases when the particle number is increased, which implies that there are no
$N$-body bound states beyond $N=N^{}_f$. Likewise, in all cases we find that, at fixed $N_f^{}$, the energy per particle heals to 
the energy of the $N/N^{}_f = 1$ case, i.e. it is dominated by the $N^{}_f$-body bound-state contribution.
To see this in more detail, we show the energy again in Fig.~\ref{Fig:2D_energy_Nf4_Nx10Diff} where we have subtracted the energy
per particle of the $N/N^{}_f = 1$ case from that of the $N/N^{}_f = 3, 5, 7$ cases. Clearly, that energy difference is
much smaller than the energy per particle of the system, which shows explicitly that the $N^{}_f$-body bound-state energy 
dominates the picture. While qualitatively this is not an unexpected result, our calculations show it in a quantitatively clear fashion.
Furthermore, this shows explicitly that no new $N$-body bound states appear beyond $N=N_f^{}$.

%

%%%%%%%%%%%%%%%%%%%%%%%
\subsection{Tan's contact}

Besides the ground-state energy, one of the most interesting quantities in many-body systems with short-range interactions
is Tan's contact~\cite{TanContact, ContactReview}. This quantity is thermodynamically conjugate to the renormalized coupling, as
shown by several authors~\cite{Werner,Valiente,ThermoContact}. Indeed, one way to find it is to determine the change in the energy with the scattering length 
(which is often referred to as the ``adiabatic theorem'').
Early on, it was shown by Tan that the contact determines the high-momentum
tail of the momentum distribution, and this was soon afterwards associated with the operator-product expansion of high-energy
physics~\cite{BraatenOPE}, and since then several authors have derived exact results in the form of sum rules for response functions
and high-energy or short-distance behavior of correlation functions.

Because our calculations used a contact interaction, the determination of the contact is essentially given by differentiation
of the ground-state energy with respect to the bare coupling. Indeed, according to the adiabatic theorem 
in 2D~\cite{Valiente, WernerCastin},

\beq
C = 2\pi \frac{\partial E_\text{GS}^{}}{\partial \ln (a_0^{}/a^{}_\text{HO})} = 
2\pi \langle \hat V \rangle \frac{\partial \ln g}{\partial \ln (a_0^{}/a^{}_\text{HO})},
\eeq
i.e. computing $C$ reduces to finding the ground-state expectation value of the potential energy operator $\hat V$ 
in the many-body problem, as the remaining factor is entirely due to two-body physics. The same is true in the present ground-state approach.

\begin{figure}[t]
\includegraphics[width=1.0\columnwidth]{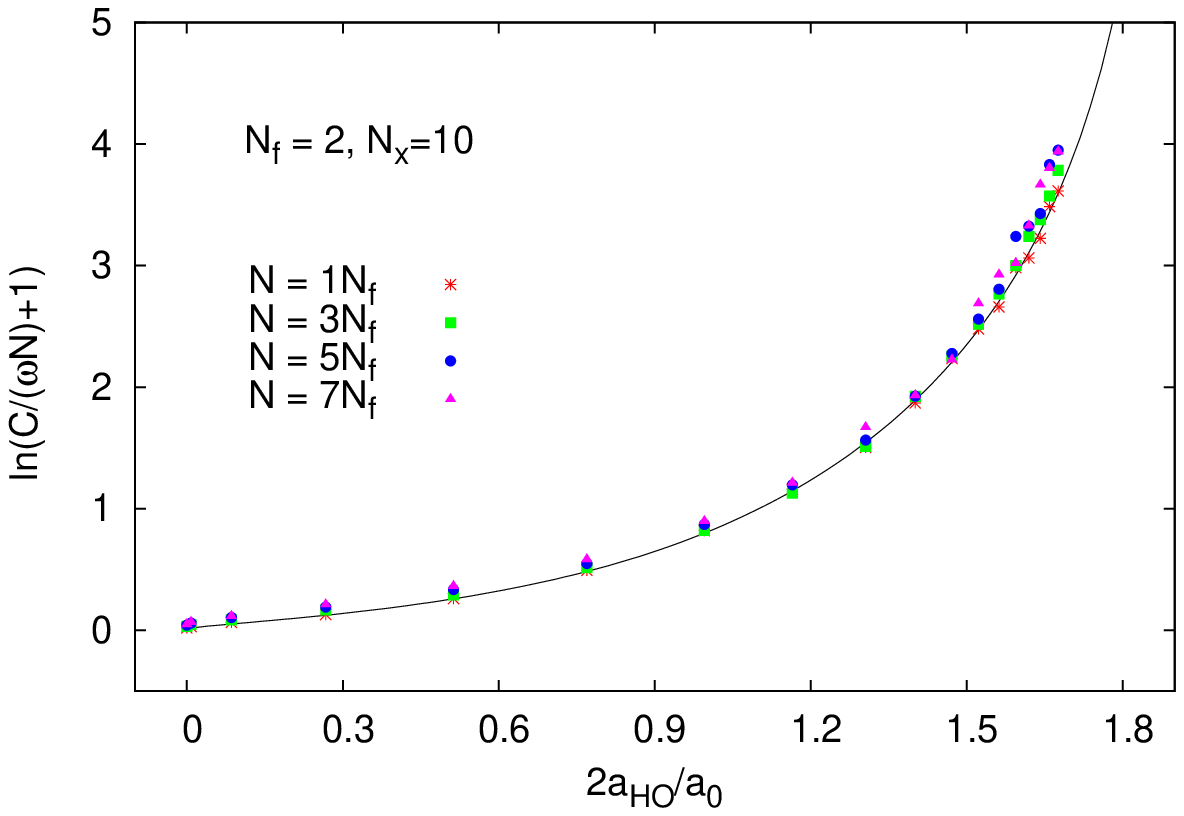}
\includegraphics[width=1.0\columnwidth]{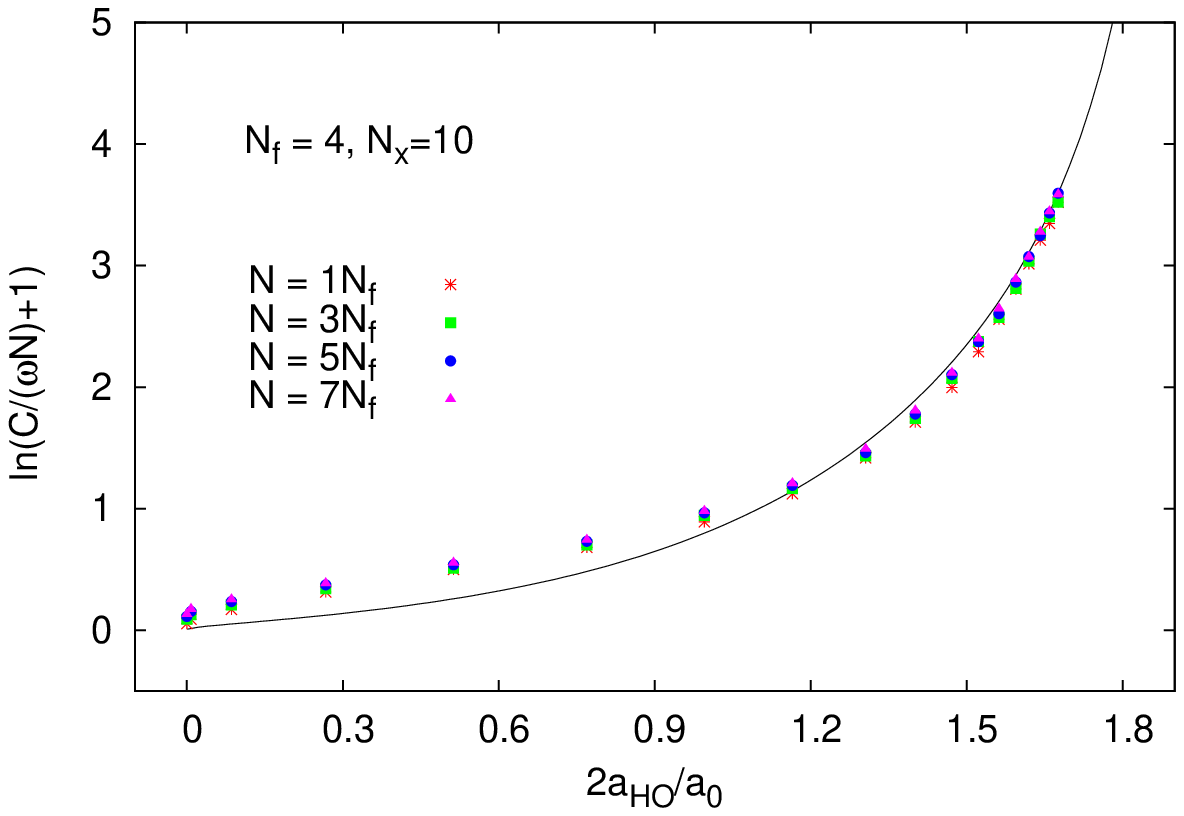}
\caption{\label{Fig:2D_contact_Nf2_Nx10}(color online)
Ground-state contact per particle of $N^{}_f=2$ (top) and $4$ (bottom) species of harmonically trapped fermions in 2D 
as a function of the coupling strength $2 a_\text{HO}/a_0$, for particle numbers $N/N^{}_f=$1, 3, 5, 7.
The solid line shows, as a reference, the two-body result for $N^{}_f=2$.
}
\end{figure}

Our results for $C$ are shown in Fig.~\ref{Fig:2D_contact_Nf2_Nx10}, where we compare the contact per particle $C/N$
for $N^{}_f=2$ and $4$ with the two-body result. While at weak couplings the $N^{}_f=4$ result is above the two-body answer,
we find that both $N^{}_f=2$ and $4$ appear to approach that answer at strong coupling. This indicates that, in absence of a
better guess, one may safely use the two-body contact in the many-body problem at strong coupling, even as $N_f^{}$
is increased.

%%%%%%%%%%%%%%%%%%%%%%%%%%%%%%%%%%%%%%%%%%%%%%%%%%%%%%%%%%%
\section{Summary and conclusions}

We used our recently proposed method of non-uniform lattice quantum Monte Carlo to analyze the 
behavior of few- to many-body systems of fermions in a two-dimensional harmonic trap.
We explored systems of $N_f^{}=2$ and $4$ flavors and up to $N/N_f^{}=7$ particles per flavor and
focused on two experimentally measurable quantities: the ground-state energy and Tan's contact.
While higher values of $N_f^{}$ are possible, we have determined that finite-size effects can be sizable 
when $N_f^{}$ is increased (although they appear to be vanishingly small for the systems
studied here). Previous work (e.g.~\cite{LiuHuDrummond} or~\cite{GharashiBlume}) studied the exact spectrum 
of the three-body problem in 2D; our work complements and extends those approaches (though restricting ourselves 
to the ground state only).
As harmonically trapped 2D systems are under intense experimental study at the moment, calculations of these basic
quantities are timely~\cite{ParishReview}. Future 2D experiments with large-$N_f^{}$ atoms can be expected, for which our
results are a prediction~\cite{CazalillaRey}.

We find that the ground-state energy per particle shows no qualitative difference for $N^{}_f=2,4$: 
it increases monotonically for all the couplings we studied when the particle number per flavor $N/N_f^{}$ is increased.
On the other hand, at fixed $N/N_f^{}$, increasing the number of flavors leads to a decrease in the energy, as expected.
In all cases, the energy is largely dominated by the $N_f^{}$-body bound state as the coupling is increased.
As the attractive interaction is thus enhanced by the addition of fermion species, a natural question is whether
new bound states arise as $N_f^{}$ is increased (i.e. beyond the one at $N/N_f^{}=1$). We find that this is not the 
case; likely the appearance of new bound states requires a finite-range interaction, as is the case in 1D
(see e.g.~\cite{BoundStates}).

%%%%%%%%%%%%%%%%%%%%%%%%%%%%%%%%%%%%%
\acknowledgements  

We acknowledge discussions with E. R. Anderson and W. J. Porter.
This work was supported by the National Natural Science Foundation of China (Grant No. 11205063).
This work was supported by the Department of Energy Computational Science Graduate Fellowship 
Program of the Office of Science and National Nuclear Security Administration in the Department of 
Energy under contract DE-FG02-97ER25308.
This material is based upon work supported by the 
National Science Foundation under Grants 
No. PHY{1306520} (Nuclear Theory Program), 
No. PHY1452635 (Computational Physics program), 
and 
No. ACI{1156614} (REU Program).

%%%%%%%%%%%%%%%%%%%%%%%%%%%%%%%%%%%%%

%%%%%%%%%%%%%%%%%%%%%%%%%%%%%%%
% Bibliography

%%%%%%%%%%%%%%%%%%%%%%%%%%%%%%%%%%%%%%%%%%%%%

\end{document}